\documentclass[preprints,article,accept,moreauthors,pdftex]{Definitions/mdpi}

\firstpage{1} 
\pubvolume{1}
\issuenum{1}
\articlenumber{0}
\pubyear{XXX}
\copyrightyear{XXX}
\datereceived{ } 
\daterevised{ } 
\dateaccepted{ } 
\datepublished{ } 
\hreflink{https://doi.org/} 
\pdfoutput=1 

\Title{Classical and Quantum Physical Reservoir Computing for Onboard Artificial Intelligence Systems: A Perspective}

\TitleCitation{Classical and Quantum Physical Reservoir Computing for Onboard Artificial Intelligence Systems: A Perspective}

\definecolor{my_green}{rgb}{0.55, 0.71, 0.0}

\newcommand{\Rey}{{\mathcal Re}}

\Author{A.~H.~Abbas \orcidA{}, Hend Abdel-Ghani\orcidC{} and Ivan S.~Maksymov\orcidB{}$^{*}$}

\address{%
Artificial Intelligence and Cyber Futures Institute, Charles Sturt University, Bathurst, NSW 2795, Australia.}

\corres{Correspondence: aborae@csu.edu.au, habdelghani@csu.edu.au, imaksymov@csu.edu.au}

\abstract{Artificial intelligence (AI) systems of autonomous systems such as drones, robots and self-driving cars may consume up to 50\% of total power available onboard, thereby limiting the vehicle's range of functions and considerably reducing the distance the vehicle can travel on a single charge. Next-generation onboard AI systems need an even higher power since they collect and process even larger amounts of data in real time. This problem cannot be solved using the traditional computing devices since they become more and more power-consuming. In this review article, we discuss the perspectives of development of onboard neuromorphic computers that mimic the operation of a biological brain using nonlinear-dynamical properties of natural physical environments surrounding autonomous vehicles. Previous research also demonstrated that quantum neuromorphic processors (QNPs) can conduct computations with the efficiency of a standard computer while consuming less than 1\% of the onboard battery power. Since QNPs is a semi-classical technology, their technical simplicity and low, compared with quantum computers, cost make them ideally suitable for application in autonomous AI system. Providing a perspective view on the future progress in unconventional physical reservoir computing and surveying the outcomes of more than 200 interdisciplinary research works, this article will be of interest to a broad readership, including both students and experts in the fields of physics, engineering, quantum technologies and computing.}


\begin{document}

\section{Introduction}
The distance between the modern world and the world depicted in science fiction has narrowed dramatically in the last two decades. However, despite the impressive progress in the development of unmanned ground vehicles (UGVs), unmanned aerial vehicles (UAVs) and underwater remotely operated vehicles (ROVs) \cite{Boy20, Dev22, Kun23}, safe and efficient fully autonomous AI-controlled machines and robots remain elusive \cite{Gia23, Zha24}.   

One of the major problems faced by the developers of autonomous platforms is high power consumption by hardware and computers \cite{Vri23, Ver23} that enable AI systems to communicate with the environment in a manner that mimics human reasoning and perception of the world \cite{Tak13}. Indeed, the rapid progress in the development of AI systems is expected to further increase the global energy consumption since approximately 1.5\,million AI server units are expected to be commissioned per year by 2027. Combined together, these devices will consume approximately 85.4\,terawatt-hours of electric power annually \cite{Vri23, Ver23}.

Several strategies have been suggested to resolve the problem of high energy consumption by AI systems. In particular, it has been demonstrated that AI systems themselves can optimise the energy consumption by forecasting supply and demand, helping prevent grid failures and increase reliability and security of energy networks that support the operation of large data centres \cite{Ghi23, Comment}. There is also a consensus in the research and engineering communities that many AI systems of the future will operate adhering to the principles of Green AI designed to reduce the climate impact of AI \cite{Ver23}.

However, even though a singe self-driving car would need to store and process small datasets compared with large amounts of data processed in data centres, the hardware of its computer may consume up to 50\% of total power available onboard of the vehicle, thus dramatically reducing the distance the vehicle can travel on a single charge \cite{Fag23, Comment1, Comment3}. Subsequently, novel approaches to the development of AI systems for autonomous vehicles need to be developed, taking into account such restrictions as the weight and size of onboard hardware and batteries \cite{Oth22, Rau24}. The latter is especially the case of specialised flying and underwater drones that must be able to reliably carry sophisticated and often power consuming equipment for long distances in complex natural environments and adversarial conditions \cite{Boy20}.
\begin{figure}[t]
\centering
\includegraphics[width=0.99\columnwidth]{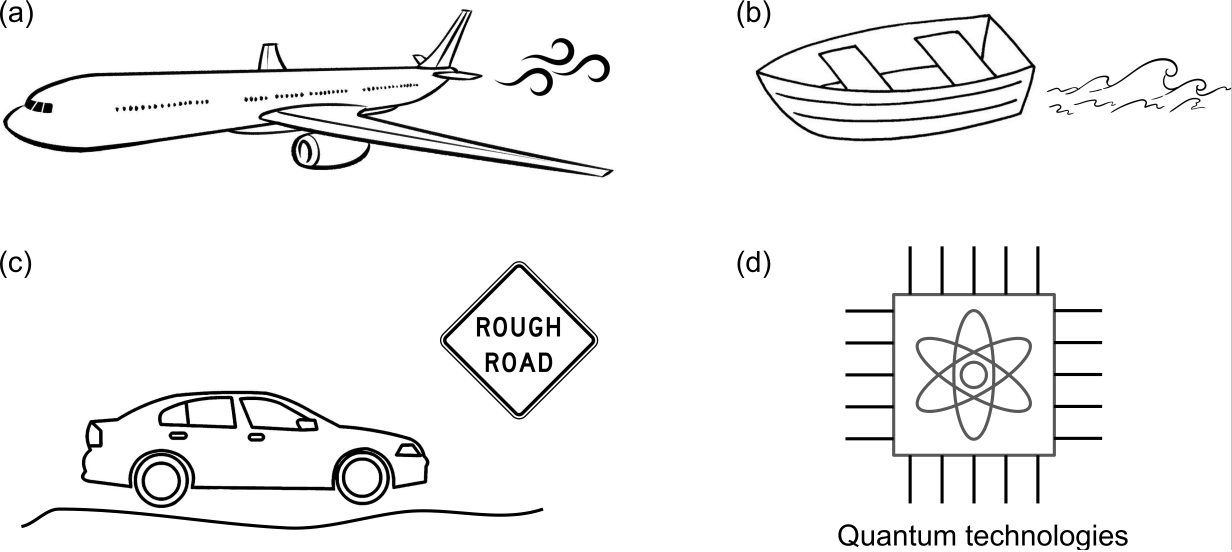}
\caption{{\bf(a--c)}~Schematic illustration of the physical processes---turbulence, water waves and vibrations caused by surface roughness---that onboard AI systems can employ as a means of energy-efficient computation. The so-envisioned AI systems can be used in UGVs, UAVs and ROVs as well as human-operated aeroplanes, boats and cars. {\bf(d)}~Identified as a separate category, quantum-mechanical physical systems can be used as onboard AI systems. Superior computational performance and low power consumption of quantum systems compared with traditional computers render them especially useful for applications in lightweight and long-range drones.}
\label{Fig1}
\end{figure}

\section{What is this review article about?}
In this article, we approach the problem of energy-efficient AI systems from a different point of view. We suggest a novel approach to the development of energy-efficient AI systems for autonomous vehicles and discuss the recent relevant academic publications. As schematically illustrated in Figure~\ref{Fig1}a--c, we propose to create AI systems that would exploit physical phenomena that either surround human-made machines, including aeroplanes, boats and cars, or are induced by their movement. For example, such AI systems would do certain computations employing turbulence in the atmosphere that forms behind an aeroplane or UAV \cite{Yan18, Sun23}. Similarly, in the case of a boat or a submersible ROV, an AI system could use water waves and other fluid-mechanical phenomena such as the formation of bubbles \cite{Yus20, Ju22} and vortices \cite{Arn15, Yu23} in water. We also demonstrate that the physical phenomena illustrated in Figure~\ref{Fig1}a--c can enable unconventional \cite{Ada17, Ada19}, neuromorphic \cite{Sha20, Mar20, Mar20_1, Sua21, Rao22, Sar22, Sch22, Kra23} and approximate \cite{Mit16, Liu20, Hen22, Ull23, Mak21_ESN} computing systems that are expected to play an increasingly important role in the development of autonomous vehicles \cite{Nom20}.

Unconventional computing is an approach to computer engineering that exploits the physical properties of mechanical \cite{Ada17}, fluid-mechanical \cite{Ada19, Mak23_review} and living \cite{Ada23, Sha24} systems to perform computations. This category also includes computer architectures that are based on electron devices that are not employed in mainstream digital and analogue computers and intended to enrich the traditional Von Neumann computer architecture and the Turing machine approaches \cite{Ada17, Ada19}.

Neuromorphic computers can be regarded as a subclass of unconventional computers that are built using some of the operating principles of a biological brain \cite{Maa02, Jae04, Luk09, Nak21}. While a neuromorphic computer may not be as universal as a traditional digital computer, it can solve certain practically important problems using readily affordable computational resources. Such an advantage originates from an inherent scalability, parallelisation and collocation of data processing and memory attainable in neuromorphic architectures \cite{Sch22}. Since a neuromorphic computer mimics the operation of a biological brain, it operates mostly when input data are available, optimising energy consumption and decreasing the cost of computations \cite{Maa02, Jae04}.

The advantages of neuromorphic computers make them ideally suitable for applications in onboard AI systems \cite{Report1}. Indeed, traditionally designed onboard AI systems have certain constraints that can introduce significant limitations to the vehicle design, creating a gap between the AI backed by high-performance computers and AI deployed in an autonomous vehicle. For example, while high-performance AI systems rely on energy-consuming multi-processor computations, systems designed for onboard use can do mostly basic calculations. Yet, even though an increase in the computational power of onboard systems is achievable by increasing the number of power supply batteries, this approach is impracticable for lightweight systems such as small and long-range drones \cite{Oku22}. On the contrary, neuromorphic computers can do approximate and energy efficient computations using limited resources \cite{Mit16, Liu20, Hen22, Ull23, Mak21_ESN}, also exploiting computational errors as a mechanism of enhancing the efficiency and additionally saving energy stored in the batteries \cite{Sch22}.

Whereas the forthcoming discussion mostly focuses on AI systems based on the paradigm of reservoir computing \cite{Luk09, Tan19, Nak20, Nak21, Cuc22}, a computational framework derived from recurrent neural network \cite{Maa02, Jae04}, the ideas and models reviewed by us can be applied to other kinds of neural network models and AI systems, including those based on deep neural network architectures \cite{Mak24_Algorithms}. In particular, a large section of this article is dedicated to quantum neuromorphic computing systems (schematically illustrated in Figure~\ref{Fig1}d) since their ability to make accurate forecasts using just a few artificial neurons \cite{Muj21, Gov21, Suz22, Gov22_1, Dud23, Got23, Llo23, Cin24}, alongside a small footprint and low energy consumption, perfectly aligns them with our vision of the application of approximate neuromorphic computing in mobile AI systems. 

The remainder of this article is organised as follows. We first review the principles of operation of reservoir computing systems, focusing on the computational algorithm known as physical reservoir computing. Then, building our discussion around the concepts illustrated in Figure~\ref{Fig1}, we provide a number of specific examples of onboard AI systems that exploit turbulence, water waves, surface roughness and quantum-mechanical effects.
\begin{figure}[t]
\centering
\includegraphics[width=0.99\columnwidth]{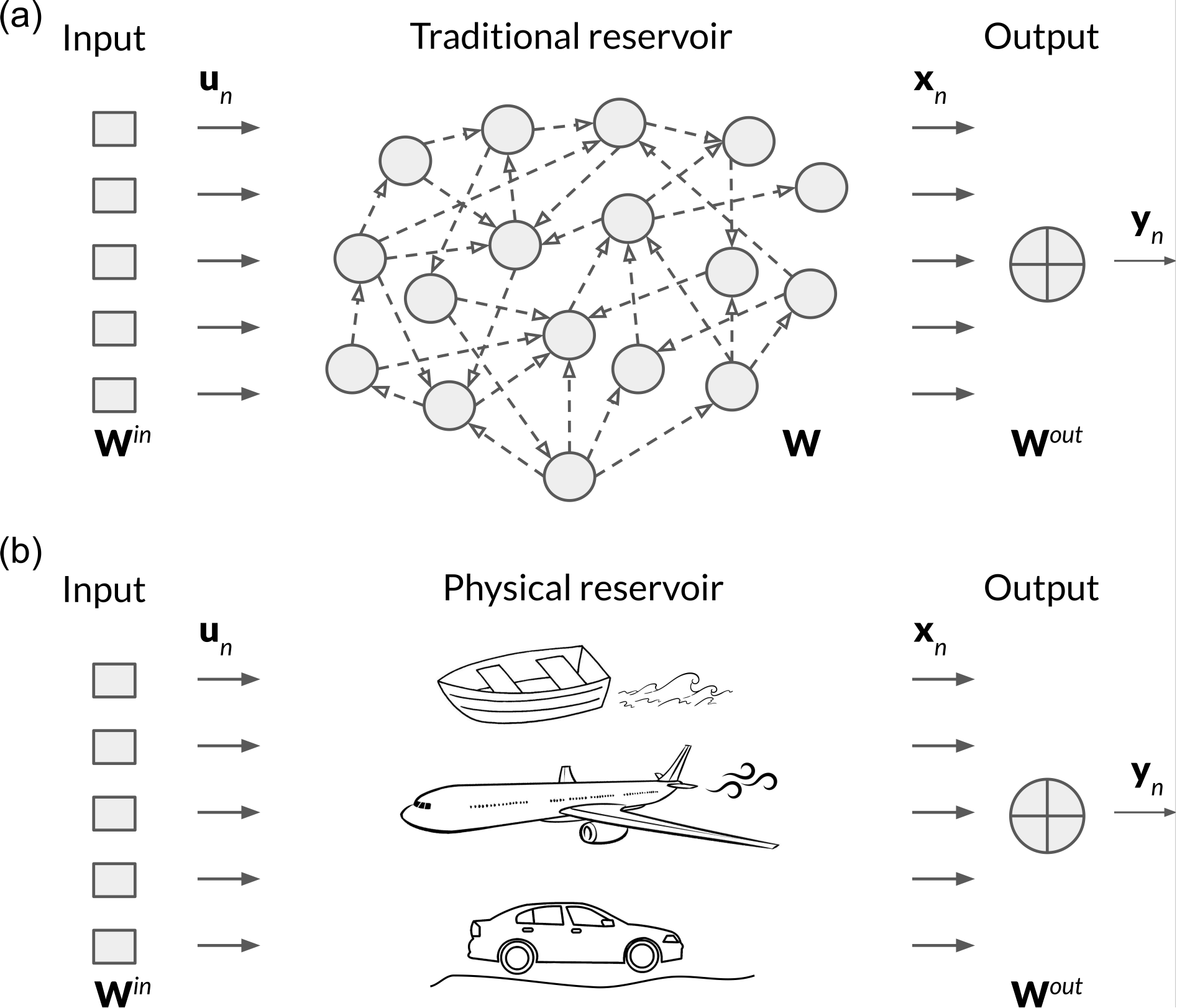}
\caption{Schematic representation of {\bf(a)}~a traditional algorithmic RC system and {\bf(b)}~a computational system with a physical reservoir constructed using the physical effects that take place in the environment that surrounds moving vehicles. The mathematical meaning of the vectors and matrices mentioned in this figure is explained in the main text.}
\label{Fig1_1}
\end{figure}

\section {Reservoir computing}
\subsection{Traditional reservoir computing approach}
Neural networks employed in AI systems usually consist of thousands or millions of interconnected processing units \cite{Vee95, Hay98, Gal07}. Each node has connections of varying strengths to other nodes in the network. The strengths of the connections is updated when the network learns to analyse large datasets.

This algorithm enables the AI system to mimic the operation of a biological brain that exploits a large and complex network of neural connections. Since the brain is also a dynamical system that exhibits complex nonlinear and sometime chaotic behaviour \cite{Mck94, Kor03} (a dynamical system is a system, of either artificial, physical or biological origin, that constantly changes its state in time \cite{Mar12}), it has been demonstrated that a neural network can be constructed using the principles of nonlinear dynamics \cite{Maa02, Jae04, Kra23}. Since in the realm of mathematics a dynamical system also employs differential equations, different kinds of nonlinear differential equations have been adopted to describe the variation of the connection strength between units of an artificial neural network \cite{Maa02, Jae04, Luk09}.

This approach to the design of artificial neural network is called reservoir computing (RC) \cite{Tan19, Nak20, Nak21, Cuc22, Mak23_review, Yan24}. In a traditional RC algorithm (Figure~\ref{Fig1_1}a), the differential equation that governs the nonlinear dynamics of the system of randomly connected nodes can be written as \cite{Jae05, Luk09, Luk12}:
\begin{eqnarray}
  {\bf x}_{n} = (1-\alpha){\bf x}_{n-1}+
  \alpha\tanh({\bf W}^{in}{\bf u}_{n}+{\bf W}{\bf x}_{n-1})\,,
  \label{eq:RC1}
\end{eqnarray}
where $n$ is the index denoting entries corresponding to equally-spaced discrete time instances $t_n$, ${\bf u}_n$ is the vector of $N_u$ input values, ${\bf x}_n$ is a vector of $N_x$ neural activations, the element-wise operator $\tanh(\cdot)$ is a sigmoid activation function \cite{Hay98}, ${\bf W}^{in}$ is the input matrix consisting of $N_x \times N_u$ randomly generated elements, ${\bf W}$ is the recurrent weight matrix containing $N_x \times N_x$ randomly generated elements and $\alpha \in (0, 1]$ is the leaking rate that controls the speed of the temporal dynamics.

Using Eq.~(\ref{eq:RC1}), one calculates the output weights ${\bf W}^{out}$ by solving a system of linear equations ${\bf Y}^{target} = {\bf W}^{out}{\bf X}$, where the state matrix ${\bf X}$ and the target matrix ${\bf Y}^{target}$ are constructed using, respectively, ${\bf x}_n$ and the vector of target outputs ${\bf y}_n^{target}$ as columns for each time instant $t_n$. Usually, the solution is obtained in the form ${\bf W}^{out} = {\bf Y}^{target} {\bf X^\top} ({\bf X}{\bf X^\top} + \beta {\bf I})^{-1}$, where ${\bf I}$ is the identity matrix, $\beta$ is a regularisation coefficient and ${\bf X^\top}$ is the transpose of ${\bf X}$ \cite{Luk12}. Then, one uses the trained set of weights ${\bf W}^{out}$ to solve Eq.~(\ref{eq:RC1}) for new input data ${\bf u}_n$ and compute the output vector ${\bf y}_n={\bf W}^{out}[1;{\bf u}_n;{\bf x}_n]$ using a constant bias and the concatenation $[{\bf u}_n;{\bf x}_n]$ \cite{Luk09, Luk12}.

RC systems have been successfully applied to solve many important problems, including the prediction of highly nonlinear and chaotic time series, modelling the variation of climate, understanding trends in financial markets and optimisation of energy generation \cite{Luk09, Luk12, Bal18, Tan19, Nak20, Cuc22, Dam22, Zha23, Mak23_review}. Importantly, similarly to the efficiency of the brain of some insects, which has a low number of neurons compared with a human brain, RC systems require just several thousands of neurons to undertake certain tasks more efficiently than a high-performance workstation computer running sophisticated software \cite{Mak21_ESN, Mak23_review}. This property is ideally suitable for the development of AI systems for mobile platforms, where the control unit must consume low power while delivering practicable machine learning, vision and sensing capabilities in a real-time regime \cite{Lee23}.   

\subsection{Physical reservoir computing}
Many physical systems and processes exhibit nonlinear dynamical behaviour and, therefore, can also be described by differential equations \cite{Mar12}. Hence, as illustrated in Figure~\ref{Fig1_1}b, it was suggested that a computationally efficient reservoir can be created using a real-life, either experimental or theoretical, nonlinear dynamical systems \cite{Luk09}. Following this idea, subsequent research works demonstrated computational reservoirs based on spintronic devices \cite{Rio19, Wat20, All23}, electronic circuits \cite{Cao22, Lia24}, photonic and opto-electronic devices \cite{Sor20, Raf20} as well as mechanical \cite{Cou17} and liquid-based \cite{Khe22, Gao22, Mak23_review, Mar23} physical systems. In the theoretical works, the demonstration of the functionality of a physical reservoir required the researchers to replace Eq.~(\ref{eq:RC1}) by a differential equation that describes the physical system of interest (e.g.,~a nonlinear oscillator \cite{Cou17, Mak21_ESN}). In turn, in the experimental works, the output of Eq.~(\ref{eq:RC1}) was effectively replaced by measured traces of temporal evolution of the physical system (e.g.,~curves of microwave power absorption caused by the excitation of spin waves in a ferromagnetic material \cite{Wat20} or optically-detected profiles of solitary waves developed on the liquid surface \cite{Mak23_EPL}). Naturally, compared with the theoretical works on physical RC systems, the experimental demonstration of physical reservoirs required an extra step that included such processing procedures as signal amplification, noise filtering and data sampling \cite{Mak23_EPL}. However, both theoretical and experimental works unanimously demonstrated substantial advantages of the physical RC systems over the algorithmic ones, including low poser consumption and high accuracy of forecasts while using a reservoir with a relatively small number of artificial neurons \cite{Mak23_review}.     
\begin{figure}[t]
\centering
\includegraphics[width=0.99\columnwidth]{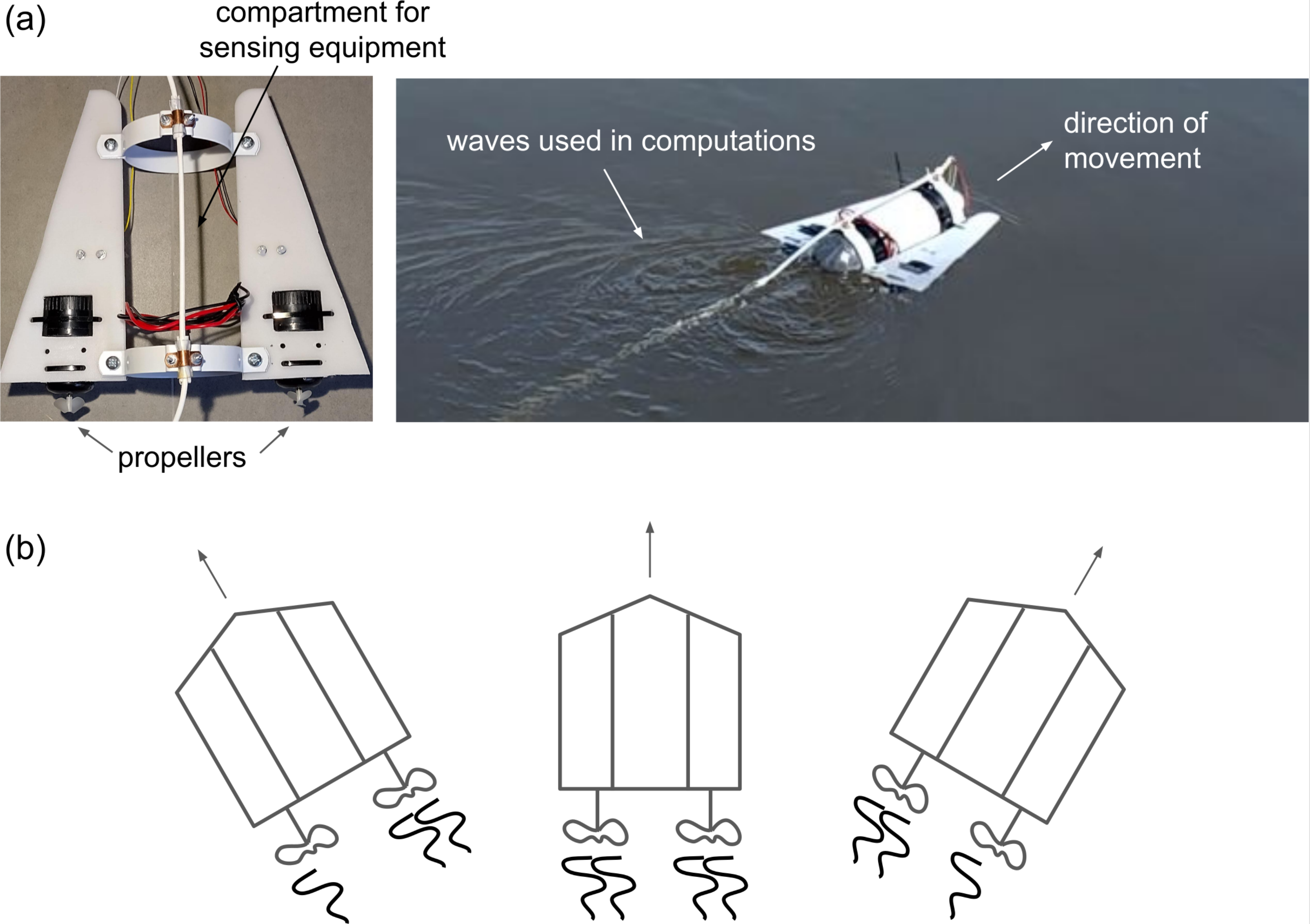}
\caption{{\bf(a)}~Photographs of the prototype of ROV designed to test AI systems that employ water waves and other disturbances caused by the motion of the drone as a means of computation. {\bf(b)}~Schematic illustration of the difference between the wave patterns produced by the ROV that has received the left, forward and right commands from the operator.}
\label{Fig2}
\end{figure}

\section{Reservoir computing using the physical properties of fluids}
\subsection{State-of-the-art}
One of the first experimental demonstrations of a physical reservoir computer was a system that exploited the ripples on the surface of water contained in a tank placed on an electric motor-driven platform \cite{Fer03} (for a review see, e.g., Refs.~\cite{Tan19, Ada19, Mak23_review}). The ripples caused by judiciously controlled vibrations were recorded using a digital camera. Then, the images were processed on a computer and then the processed data were used as the input of the reservoir computing algorithm.

Although that pioneering work clearly demonstrated the potential of liquids to perform RC calculations, for about two decades it was considered to be mostly of fundamental interest to researchers working on physics-inspired AI systems \cite{Nak15, Nak20, Tan19}. A similar idea was exploited in the recent theoretical \cite{Mar20, Mar23} and experimental works \cite{Mak23_EPL}, where it was suggested that the ripples created by motors can be replaced by the dynamical properties of solitary waves---nonlinear, self-reinforcing and localised wave packets that can move for long distances without changing their shape \cite{Rem94}.

Importantly, those novel works demonstrated the ability of water-based RC system to do complex calculations relying only on low computational power microcontrollers (e.g.,~Arduino models \cite{Mak24_dynamics}) that can, in principle, continuously operate for several months without the need to recharge the battery. Therefore, such systems are especially attractive for application in drones, robots and other autonomous platforms. 
\begin{figure}[t]
\centering
\includegraphics[width=0.9\columnwidth]{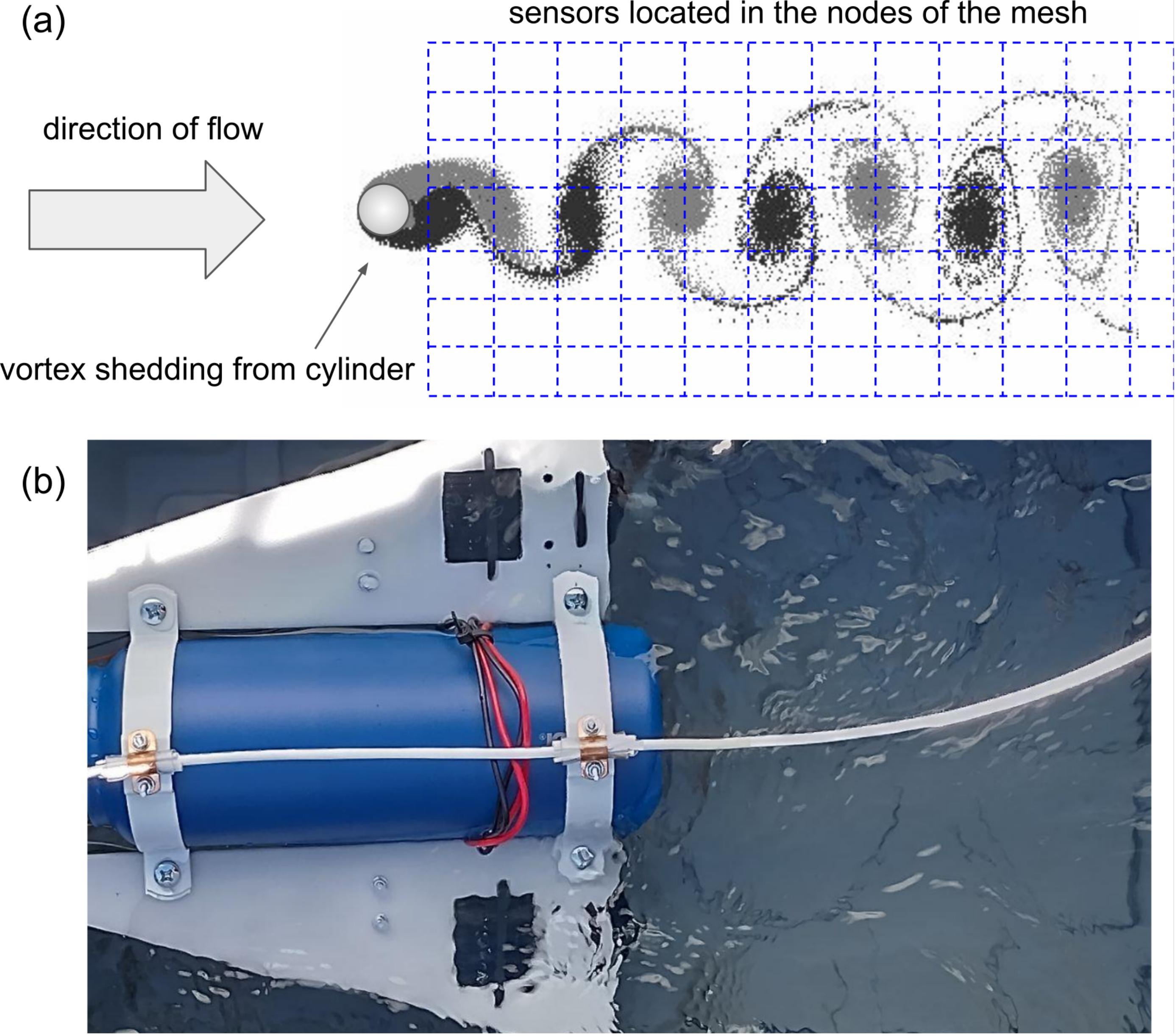}
\caption{{\bf(a)}~Illustration of the vortex shedding taking place when a fluid such as air or water flows past a cylinder. As theoretically shown in Ref.~\cite{Got21}, modulating the flow velocity and monitoring the vortex dynamics using a set of virtual sensors one can create an efficient physical RC system. An experimental implementation of this computational approach was discussed in Ref.~\cite{Vin23_1}. {\bf(b)}~Photograph of vortices and other water flow effects created by the ROV in a lab setting.}
\label{Fig3}
\end{figure}

\subsection{Towards Reservoir Computing with Water Waves Created by an ROV}
Figure~\ref{Fig2}a shows the photographs of an experimental ROV designed to test different kinds of autonomous AI systems. The hull of ROV consists of two wing-like fins that hold two electric motors that rotate the propellers: one in the counterclockwise direction and another one in the clockwise direction \cite{Agu19}. The fins are connected to a cylindrical frame that is also used as a compartment where sensing equipment, including lasers and photodetectors, can be placed. As with many commercial ROV designs \cite{Boh97}, the movement of the experimental ROV is controlled via an electric cable.

When both propellers rotate at the same speed, the ROV moves forward (Figure~\ref{Fig2}b). However, the ROV will turn left (right) when the speed of the left (right) motor is reduced. Thus, a moving ROV creates a wake wave whose pattern depends on the relative speed of the two motors. The temporal dynamics of the wave pattern can be sensed using two laser-photodiode pairs that measures the light reflected from the waves created by the left and right propellers (a similar detection mechanism was used in Ref.~\cite{Mak24_dynamics}). The resulting optical signals can then be converted in electric signals that are transmitted to the control unit via the cable. The waveform of the so-generated electric signals will be a function of time and it will correlate with the commands (left, right, forward and so on) given to the ROV by the operator. Indeed, as schematically shown in Figure~\ref{Fig2}b, when the ROV moves forward the waves created by both propellers are approximately the same. However, the wave patterns change when the ROV turns left or right.

Similarly to the seminar work Ref.~\cite{Fer03}, the differences in the wave pattern of the moving ROV can be employed to perform reservoir computations. Such computations can help predict an optimal trajectory of the ROV based on the previous inputs of the human operator and taking into account the environmental factors, including the speed of wind, temperature and salinity of water as well as the presence of obstacles such as rocks and debris \cite{Yan23, Per24}. Since the ROV creates the waves that are used to predict the trajectory, the on-board reservoir computer does not consume significant energy apart from the need to power the sensors, thereby satisfying the requirements for autonomous AI systems \cite{Hen22}.

\subsection{Physical Reservoir Computing using Fluid Flow Disturbances}
The research work involving the ROV shown in Figure~\ref{Fig2} is still in early development. However, at the time of writing, to the best of our knowledge it is the only experimental attempt to build a prototype of a water drone that employs physical processes in its environment as a means of computation (a UGV designed using similar principles is discussed in Section~\ref{UGV}). Nevertheless, similar ideas have been explored theoretically and also tested in laboratory settings \cite{Got21, Mak21_ESN, Vin23, Vin23_1, Mak23_EPL, Mak24_dynamics}. Although the results presented in the cited papers do not involve any autonomous vehicle, in the following we will analyse them in the context of the ROV model discussed above. 

Before discussing those results, we also highlight the proposals of physical RC systems designed to predict the trajectory of a flying drone \cite{Var15, Per24}. Although the idea of using physical processes taking place in the surrounding environment of the drone were not expressed in Ref.~\cite{Var15, Per24}, the approach presented there can be extended to implement the concepts illustrated in Figure~\ref{Fig1}a,~b. In this context, we remind that the discipline of fluid dynamics is concerned with the flow of both liquids and gases \cite{Sea11}. This means that AI systems designed for flying and underwater drones can, in principle, exploit the same physical processes. In fact, as mentioned above, both moving airborne and underwater vehicles create vortices as they move in the atmosphere \cite{Sab09, Hru22, Nat22} and water \cite{Yu23}.

Vortices are ubiquitous in nature (e.g., they can be observed in whirlpools, smoke rings and winds surrounding tropical cyclones and tornados) and they exhibit interesting nonlinear dynamical that has been the subject of fundamental and applied research \cite{Vor_book, Bil91, Wil96, Wil04}. For instance, it is well-established that when fluid flows around a cylindrical object, the physical effects known as von K{\'a}rm{\'a}n vortex street can be observed in wide a range of flow velocities \cite{Wil96, Vor_book}. The shedding of such vortices imparts a periodic force on the object. In many situations, this force is not significant enough to accelerate the object. However, in some practical cases the object can vibrate about a fixed position, undergoing harmonic motion. Yet, when the frequency of the periodic driving force matches the natural frequency of the oscillation of the object, the resonance processes come into play and the amplitude of the oscillations can increase dramatically \cite{Wil04, Bil91}.
\begin{figure}[t]
\centering
\includegraphics[width=0.99\columnwidth]{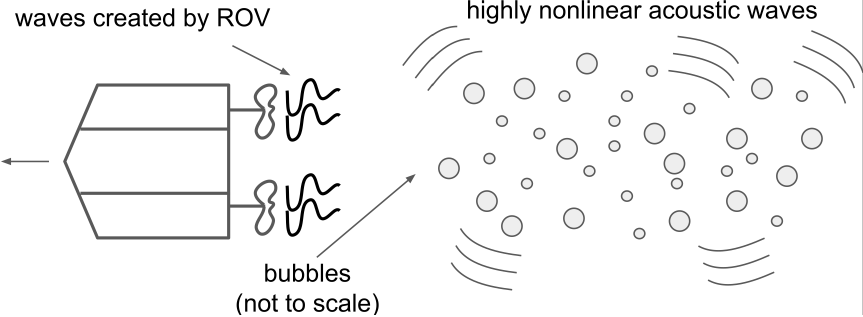}
\caption{Sketch of an ROV, bubbles created by its propellers and highly nonlinear acoustic waves emitted by them. As discussed in the main text, the nonlinear dynamics of these physical processes opens up novel opportunities for physical reservoir computing.}
\label{Fig4}
\end{figure}

Therefore, it has been theoretically demonstrated that the physical properties of vortices can be used in reservoir computing \cite{Got21, Vin23}. In Ref.~\cite{Got21}, the authors conducted a rigorous numerical analysis of a vortex-based RC system based on a von K{\'a}rm{\'a}n vortex street (Figure~\ref{Fig3}a). A periodic pattern of numerical sensors located across the computational domain was used to monitor the nonlinear dynamics and collect data for further processing following the traditional RC algorithm. The flow of fluid was used as the input. For example, to create an input that corresponds to a signal that varies in time, the velocity of the flow was modulated such that it follows the time-varying shape of the signal (which can readily be done using an electric pump \cite{Mak23_EPL, Mak24_dynamics}).     

The dynamics of the resulting computational reservoir depends on the value of the Reynolds number $\Rey$. Subsequently, different operating regimes were tested using inputs corresponding to the values of $\Rey$ below and above the threshold of the formation of a von K{\'a}rm{\'a}n vortex street \cite{Got21}. Those tests confirmed a high memory capacity of the reservoir and its ability to learn from input data and generalise them. Further tests also revealed the ability of the reservoir to make accurate predictions of time series datasets.
\begin{figure}[t]
\centering
\includegraphics[width=0.99\columnwidth]{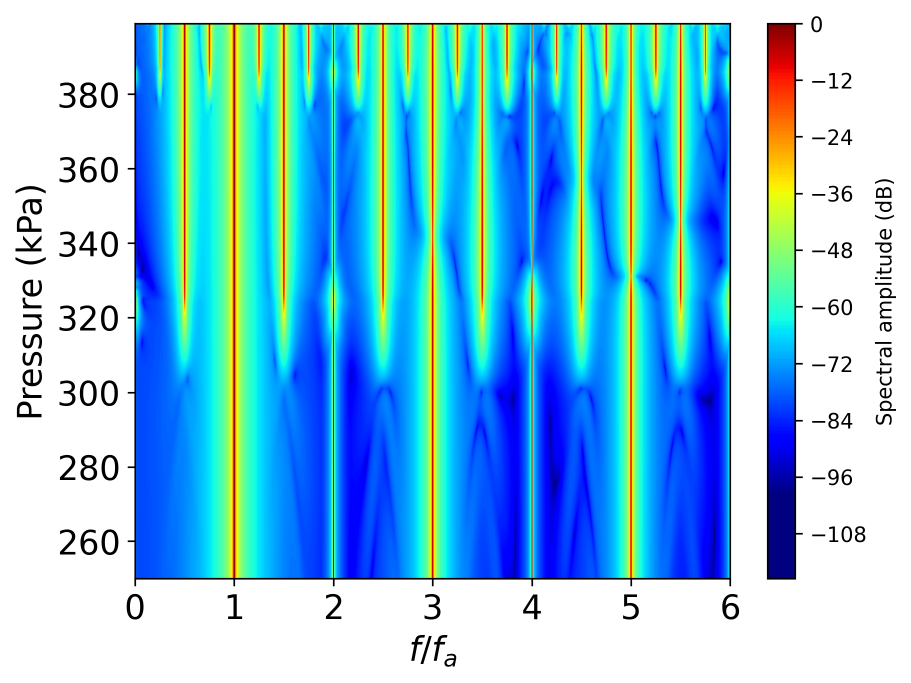}
\caption{False-colour map composed of a large number of theoretical frequency spectra of a single gas bubble trapped in the bulk of water and excited with a sinusoidal acoustic pressure wave. The $x$-axis corresponds to the normalised frequency $f/f_a$, where $f_a$ is the frequency of the incident acoustic wave. The $y$-axis denotes the peak pressure of the incident acoustic wave. Using this figure, we can see the evolution of the frequecny spectrum of the bubble as a function of the peak pressure of the incident wave. For example, at 260\,kPa we can identify the frequency peaks with the frequencies $f/f_a = 1, 2, 3$ and so on. At 320\,kPa, we observe the appearance of the subharmonic peak at $f/f_a = \frac{1}{2}$ and its ultraharmonic components $f/f_a = \frac{3}{2}$,~$\frac{5}{2}$ and so on.}
\label{Fig5}
\end{figure}

We note that in an experimental attempt to test the theoretical vortex-based RC system \cite{Vin23_1} the virtual sensors shown in Figure~\ref{Fig3} can be replaced by real ones. Alternatively, one can use a digital camera to film the vortex and then process different pixels of the individual frames extracted from the video file \cite{Mak23_EPL, Mar23, Vin23_1, Mak24_dynamics}. It is also noteworthy that propellers of the ROV can also create vortices \cite{Asn21, Yu23} that can be used for reservoir computing purposes (Figure~\ref{Fig3}b). 

\subsection{Acoustic-based Reservoir Computing}
Another promising approach to physical reservoir computing employs acoustic waves, vibrations and adjacent physical processes \cite{Mef23, Yar23, Pha23}. Although high-frequency (MHz-range) acoustic waves have been used in the cited papers, the ideas presented in those works can be implemented using the acoustic phenomena observed in a wide range of frequencies. For example, the temporal dynamics of vortices and other disturbances created by ROVs (Figure~\ref{Fig3}b) has a spectral signature in the frequency range that spans from several tens of Hz to several hundreds of kHz. Such acoustic signals can be detected using hydrophones, sonar technologies and other well-established acoustic location techniques \cite{Wil04_fdtd, Rub99, Man99, Dig16} and then processed using an RC algorithm \cite{Ona22}.

As with the sound radiated by a moving ship \cite{Lid15}, ROVs and similar autonomous vehicles can produce a tonal (related to the blade pass frequency) acoustic disturbance and broadband noise associated with the presence of unsteadiness in the flow. The tonal disturbances can be further categorised into the contributions related to such technical parameters of the propellers as the blade thickness parameter and blade loading \cite{Lid15}. These acoustic processes also exhibit significant nonlinear effects \cite{Mar70} that can be exploited in a computational reservoir \cite{Tan19, Nak20, Mak23_review}.

Moreover, as the propeller rotates it pushes the ROV through the water, causing a positive pressure on the face of the blade and a negative pressure on its back. The negative pressure causes any gas in solution in the water to evolve into bubbles \cite{Ple49, Bre95, Lau10, Mak22_Bio, Mak22_combs}. These bubbles collapse via the process called cavitation, causing hammer-like impact loads on the blades and damaging their surface \cite{Lau10, Mak22_Bio, Mak22_combs}. The cavitation also causes significant acoustic noise that originates both from oscillations and collapse of bubbles \cite{Lau10} and the formation of vortices \cite{Arn15, Yan23}. This physical picture is sketched in Figure~\ref{Fig4}.

In particular, the underwater acoustic noise is associated with the highly-nonlinear oscillation of the bubble volumes that typically occur in the range of frequencies from several hundred Hz to approximately 40\,kHz \cite{Bre95, Lau99, Lau10, Mak19} (microscopic bubbles oscillate at higher frequencies \cite{Che09, Sus12, Dza13}; however, the physics of their acoustic response remains essentially the same). Indeed, considering an idealised scenario of a sinusoidal acoustic pressure wave, when a wave moves through water, its initial waveform changes so that its initial monochromatic spectrum acquires higher harmonic frequencies \cite{Mak19}. The more nonlinear the medium in which the wave propagates, the stronger the enrichment of the spectrum with the peaks corresponding to harmonics.

The degree of acoustic nonlinearity can be characterised by the acoustic parameter $\beta=B/A$, which is the ratio of coefficients $B$ and $A$ of quadratic and linear terms in the Taylor series expansion of the equation of state of the medium (see \cite{Mak19} and references therein). The larger the value of $\beta$, the more nonlinear the medium, the stronger a distortion of the acoustic spectrum from the initial monochromatic state. For instance, water with $\beta=3.5$ is more acoustically nonlinear than air with $\beta=0.7$ (we note that the degree of nonlinearity is considered to be moderate in both media).

However, when air bubbles are present in water, the value of $\beta$ increases to around 5000 \cite{Mak19}. This is because liquids are dense and have little free space between molecules, which explains their low compressibility. In contrast, gases are easily compressible. When an acoustic pressure wave propagating in water reaches a bubble, due to the high compressibility of the gas trapped in it, its volume changes dramatically, causing substantial local acoustic wavefront deformations that result in strong variation of the initial acoustic spectrum.

The evolution of the acoustic spectrum of an oscillating bubble trapped in the bulk of water is illustrated in Figure~\ref{Fig5} (for the computational details and model parameters see, e.g., \cite{Sus12, Tony21, Mak22_Bio, Mak22_combs}). The units of the $x$-axis of this figure correspond to the normalised frequency $f/f_a$, where $f_a$ is the frequency of the incident sinusoidal acoustic pressure waves. The $y$-axis corresponds to the peak pressure of the incident wave (in kPa\,units) but the false colour encodes the amplitude (in dB\,units) of the acoustic pressure scattered by the bubbles. The bright traces with the amplitude of approximately 0\,dB correspond to the frequency peaks in the spectra of the bubble forced at one particular value of the peak pressure of the incident wave. We can see that at a relatively low pressure of the incident wave the spectrum contains the frequency peaks at the normalised frequencies $f/f_a = 1, 2, 3$ and so forth. However, an increase in the pressure results in the generation of the subharmonic frequency $f/f_a = \frac{1}{2}$ and its ultraharmonic components $f/f_a = \frac{3}{2}$,~$\frac{5}{2}$ and so on. Further increase in the peak pressure of the incident waves leads to a cascaded generation of subharmonics frequency peaks, resulting in a comb-like spectrum \cite{Mak22_combs, Tony21}.  

The nonlinear response of a cluster of oscillating bubbles trapped in water was used to create a computational reservoir in theoretical work Ref.~\cite{Mak21_ESN}. Each bubble in the cluster oscillates at a certain frequency when the entire cluster is irradiated with an acoustic wave. The oscillation frequency of each bubble in the cluster depends on the equilibrium radius of the bubble and the strength of its interaction with the other bubbles in the cluster. Importantly, the cluster maintains its structural stability (i.e. the bubble do not merge) when the pressure of the incident acoustic wave remains relatively low \cite{Tony21}.

Thus, when the temporal profile of the incident pressure wave is modulated to encode the input data (e.g., a time series that needs to be learnt and then forecast by the RC system), the cluster of bubbles acts as a network of artificial neurons. Since each oscillating bubble emits sound waves, these waves can be detected using either a hydrophone or a laser beam \cite{Mak22_Bio, Mak22_combs} and then process following to the traditional RC algorithm. As demonstrated in Ref.~\cite{Mak21_ESN}, this computational procedure enables the RC system to predict the future evolution of highly nonlinear and complex time series with the efficiency of the traditional RC algorithm while using low energy consuming computational resources.
\begin{figure}[t]
\centering
\includegraphics[width=0.99\columnwidth]{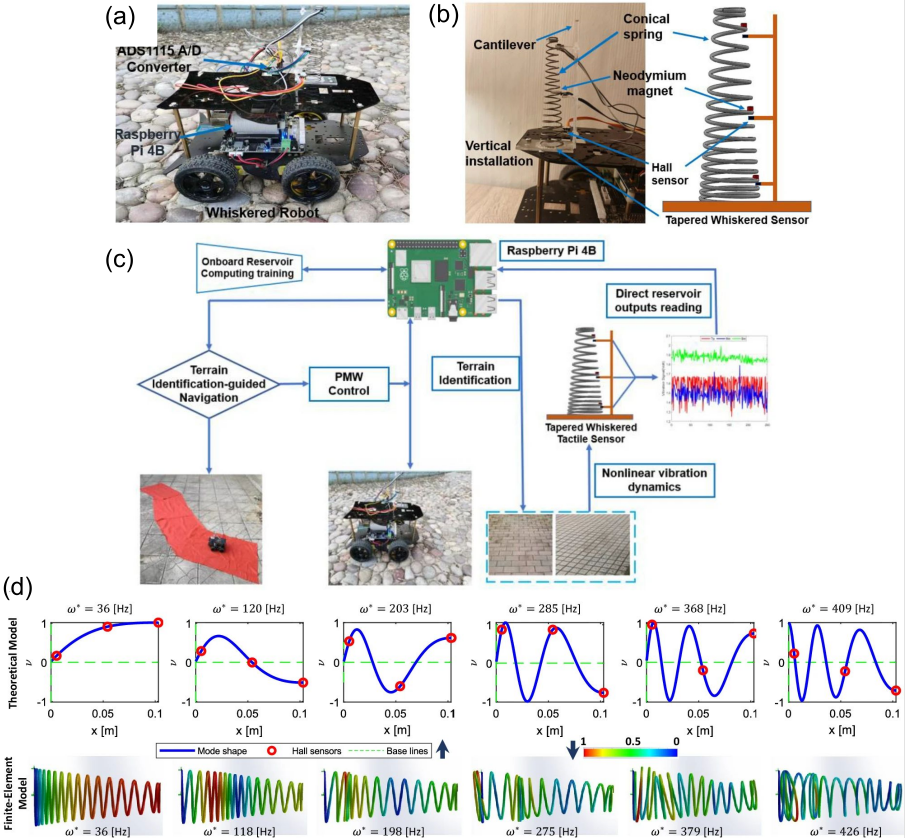}
\caption{{\bf(a)}~Photograph of the UGV controlled by a physical RC system that employs the road roughness and vibrations of a whiskered sensor as a means of computation. {\bf(b)}~Whiskered sensor and the detectors used to monitor the nonlinear vibrations of the spring caused by the rough terrain. {\bf(c)}~Schematic diagram of the onboard reservoir computer that exploits the nonlinear dynamics of the vibrating spring and processes data using a microcontroller. Identifying the texture of the terrain, the microcontroller computes as optimal trajectory of the vehicle. {\bf(d)}~Axial displacement and natural angular frequency of the vibrational modes of the tapered spring obtained using a theoretical model (top row) and rigorous numerical simulations by means of a finite-elements method. Adapted from Ref.~\cite{Yu23_1} under the terms of a Creative Commons Attribution 4.0~International License.}
\label{Fig6}
\end{figure}

Subsequently, it is plausible that bubbles created by a moving ROV (Figure~\ref{Fig4}) can be used to construct an onboard RC system. The input of such a reservoir will be the local pressure variations caused by the propellers of the ROV. It can be shown that these variations correlate with the control signals received by the ROV from the operator and/or its onboard control unit \cite{Pat20, Su24}. While the implementation of such a computational scheme requires resolving several technological problems, it enables the researchers to access a rich spectrum of fascinating nonlinear effects associated with oscillating bubbles \cite{Pro74, Kel80, Lau10, Sus12, Dza13, Mak19}. Interestingly enough, as demonstrated in the following section, a similar, from the point of view of nonlinear dynamics, idea was proposed in the domain of autonomous ground vehicles.

\section{Physical Reservoir Computing for UGVs\label{UGV}}
The preceding discussion of the fluid flow disturbances and their application in reservoir computing should also be applicable to cars \cite{Pau19, Nak22_1}, bicycles \cite{Mia20} and other road vehicles \cite{Gar22} that create turbulence and give rise to other physical effects that can be used in reservoir computing. However, the physical contact of road vehicles with the ground often results in unique nonlinear dynamical processes that can be employed in a physical reservoir computer, as schematically illustrated in Figure~\ref{Fig1}c.  

To date, different AI-based approaches to terrain identification have been developed since this functionality is essential for autonomous vehicles and robots that operate in extreme and unstructured environments \cite{Tra11, Ots16, Chr16, Val17}. Identifying the terrain surface and perceiving its texture, UGVs and robots can dynamically adjust their initial trajectory, achieving safer and more efficient navigation with the help of neural network models. Similarly to flying and underwater drones, UGVs and robots must be able to perceive the surrounding environment with high accuracy, consuming low power and using a low computational onboard computer that can process data in a real-time regime.

A range of physical sensing systems have been developed to fulfil the aforementioned technical specifications. For example, computer vision techniques have been used to enable robots to recognise different terrain textures from a longer distance to modify the route and avoid obstacles \cite{San11, Nav19}. LIDAR (Light Detection and Ranging) and hybrid optical-machine vision systems have also been proposed in Refs.~\cite{Kon09, Zho12}. Nevertheless, despite the encouraging results demonstrated in those works, the accuracy of terrain monitoring enabled by visual methods is adversely affected by variations in light intensity, limited visibility conditions caused, for example, by smoke, and weather-related factors such as rain, snow, fog and fallen leaves.

Yet, the accuracy of many optical sensing methods deteriorates due to vibrations. Of course, one can use different signal processing techniques to mitigate the adverse effect of vibrations \cite{Eng23}. However, the implementation of such approaches increases the complexity of software that controls the vehicle, which, in turn, requires a more powerful onboard computer that consumes the energy stored in the vehicle's batteries.

On the other hand, it has been suggested that acoustic effects \cite{Chr16} and vibrations \cite{Bro05} can be used as a means of terrain classification. Indeed, it is plausible to assume that vibrations of certain structural elements of a vehicle will vibrate following the roughness of the terrain, thereby providing a valuable information from which the structure of the terrain can be deduced. However, the extraction of vibration profiles and integration of those data with the existing machine learning techniques have proven technically difficult and computationally demanding \cite{Gig11, Val17}.

An innovative approach to the solution of this problem has been proposed in Ref.~\cite{Yu23_1}, where it was demonstrated that the nonlinear dynamical response of the vehicle's mechanical features can be used as the means of computations. That is, instead of trying to either mitigate the effect of vibration or process vibration-induced signals using an onboard AI system, it was proposed that vibrations could serve as the input of a reservoir computer (Figure~\ref{Fig6}a). 

Previous research demonstrated that a structural part of a soft robot can be used as a computational reservoir \cite{Nak13}. Following that idea, in Ref.~\cite{Yu23_1} used a whisker sensor that mimics the tactile sensing capabilities displayed by various living organisms, particularly insects and mammals \cite{Wan23}. The particular whisker sensor geometry used in Ref.~\cite{Yu23_1} is a tapered spring (Figure~\ref{Fig6}a,~b) that vibrates as the vehicle moves through a rough terrain, producing nonlinear signals that are detected using three Hall sensors and permanent magnets (Figure~\ref{Fig6}b,~c). The so-measured signals are processes using a low-power consuming onboard microcontroller that implements the basic reservoir computing functionality (Figure~\ref{Fig6}c). Identifying the texture of the terrain, the microcontroller steers the vehicle in the desired direction, varying the electric power delivered to the motors. Essentially, this control scheme is similar to the ideas expressed in the preceding sections of this articles:~onboard sensors of a moving vehicle detect nonlinear-dynamical processes occurring in the outer environment and the signals produced by them are employed as a means of reservoir computation aimed to identify an optimal route as well as to perform other navigation and energy consumption optimisation functions.

Importantly, it has been demonstrated that the nonlinear dynamics of the whiskered sensor can be captured using just three sensors places along the tapered spring (Figure~\ref{Fig6}b). The optimal location of the sensors was computed based on the information about the vibrational modes of the spring obtained using a theoretical model and rigorous numerical simulations (Figure~\ref{Fig6}d). It was shown that the sensors located closer to the base of the tapered spring follows the vibration caused by the rough terrain. However, the second and third sensors located in the middle of the spring and close to its tip, respectively, mostly detect the higher-order vibrational modes. Such a sensor arrangement was found to be optimal for the construction of an  efficient computational reservoir. Interestingly, this observation agrees with the previous theoretical demonstration of the RC system based on nonlinear oscillations of bubbles trapped in water \cite{Mak21_ESN}, where it sufficed to take into account the fundamental mode and a few higher-order nonlinear acoustic oscillation modes of the bubble to perform complex forecasting tasks.            

\section{Adjacent technologies for onboard reservoir computations}
Thus, mechanical whiskered sensors and adjacent sensor technologies open up exciting opportunities for the detection of nonlinear variations in the environment and their application in reservoir computing. In fact, different kinds of whiskered sensor---optical, magnetic, resistive, capacitive and piezoelectric---have also been used in underwater ROVs \cite{Wan23}, which means that not only UGVs but the other types of autonomous vehicles can employ the physics of whiskered sensors for computational purposes.

However, every technology has technical and fundamental limitations. For example, magnetic whiskered sensor can be sensitive to temperature fluctuations, which may renders them inefficient in certain practical situations \cite{Wan23}. Moreover, from the fundamental point of view, the operation of a whiskered sensor requires the application of relatively strong magnetic fields \cite{Tri19}. Subsequently, strictly speaking, such sensors are not suitable for the detection of small magnetic field variations in the surrounding environment.

Magnetic phenomena have been widely used in the domain of reservoir computing, where a number of magnonic \cite{Fur18, Wat20} and spintronic \cite{Tan22, Vid23, Edw23} RC systems have been demonstrated. Therefore, it is conceivable that a sensor located onboard of an autonomous vehicle can take a series of measurements to determine the magnetic field surrounding the vehicle, which can then be compared to the already known magnetic field maps to provide valuable information about the vehicle's location and serve as a means for reservoir computing. 

While certain ultra-sensitive measurement techniques developed in the field of microwave magnetism \cite{Iva98, Mak15} could be used to implementation this idea in practice, mostly quantum sensors that can measure small magnetic field variations with high accuracy \cite{Deg17} have the potential to serve as an onboard quantum reservoir computer (Figure~\ref{Fig1}d). Indeed, a number of onboard quantum sensors have been developed and tested both in lab and real-life environments \cite{Jes16, Tem22, Tas24}. Therefore, in the following section, we discuss the emergent quantum reservoir computing architectures that can be used to conduct complex real-time calculations onboard of autonomous vehicles.   

\section{{Quantum Reservoir Computing}\label{Quantum}}
Developing a functional quantum computer presents numerous challenges, primarily due to the fragile nature of quantum bits (qubits). Qubits are highly susceptible to decoherence and various forms of quantum noise, which complicates the engineering of a high-fidelity processor capable of executing quantum algorithms across an exponentially large computational space \cite{Pre18, Aru19}. Subsequently, robust mechanisms for precise error correction are essential for maintaining computational accuracy. Yet, these mechanisms must be suitable for applications at extremely low temperatures.

To resolve this problem, alternative avenues for harnessing the power of quantum physics for computational tasks have been explored. Quantum reservoir computing (QRC) is one of these approaches \cite{Nak16}. QRC systems mimic the computational capabilities of the human brain by using quantum dynamics, which is an innovative approach that holds a tremendous potential to advance the research on quantum AI and neuromorphic computing \cite{Sch22}. One particular advantage of QRC is low training cost and fast learning capabilities, which makes it especially efficient for processing complex data \cite{Abb24}.

Essentially, QRC is a quantum dynamical system whose quantum states are represented by the density matrices $ \rho_k \in T(\mathcal{H}) $, where $ T(\mathcal{H}) $ denotes the space of Hermitian, positive semi-definite, trace-one operators. Since quantum dynamics is inherently linear, the achievement of nonlinearity in QRCs requires one to careful choose the quantum observable that can produce a nonlinear input-output transformation. An efficient computational reservoir should also fulfil the echo state and fading memory conditions \cite{Jae04, Luk09, Tan19, Nak20}, which implies that its temporal dynamics must be dissipative and independent of the initial conditions \cite{Par24, San24}.

In QRC, Fock states of the quantum system serve as the neural activations of the reservoir \cite{Mar20_abbas, Gov21, Abb24}. Created by a tensor product of quantum subsystems, Fock states enable a unique way to model the activity of neurons. In such a quantum system, external signals prompt the quantum states to evolve dynamically over time. The evolution of these quantum states is governed by a unitary operator that is determined by the Hamiltonian of the system, a fundamental concept in quantum physics that describes the total energy of the system and dictates how the quantum states change over time \cite{Gri04, Nie02}. As the system evolves, it processes the input signals through its inherent quantum dynamics.

The output from each neuron (i.e.~the Fock state) is then collected and processed using a linear function. This function aggregates the contributions from individual neurons, providing a combined output with different weights that represent the response of the system to the input signals. The weights associated with each neuron’s output are then trained using linear regression techniques, a statistical method that adjusts the weights to minimise the error between the predicted and actual outputs. By training the weights through linear regression, the QRC network can be fine-tuned to improve its ability to process and interpret input signals, leading to more accurate and efficient performance.

A variety of technological platforms were proposed to function as QRCs, including trapped ions \cite{zha17, pin21}, nuclear magnetic resonance (NMR) in molecules \cite{Neg18}, quantum circuits \cite{chen20, das22} and photonic devices \cite{cai17, Nok18}. More recently, novel QRC systems have emerged, including arrays of Rydberg atoms \cite{Bra22} and Josephson mixers \cite{Dud23}. In the theoretical domain, models involving the quantum master equations have been developed to describe quantum spins with controlled losses, enabling the studies of coherence, scalability and controllability of the system \cite{Dud23, Abb24}.

Another major advantage of QRC is a large number of degrees of freedom available in small quantum systems. To optimise the extraction of information from these degrees of freedom, techniques like temporal multiplexing, which improves the performance of QRC models, and spatial multiplexing, which simultaneously uses multiple reservoir layers, have been proposed \cite{Nak19}.

In the following subsections, we will survey the main categories of QRC systems, exploring their key characteristics and discussing their computational capabilities.

\subsection{Spin-Network Based Reservoir}
A spin-network system is a promising platform for implementing QRC algorithms. Indeed, extensive research has delved into the dynamics of spin chain networks, showcasing their potential for complex computational tasks \cite{Mar05, Tse11}. QRC systems utilising a quantum network of randomly coupled spins Refs.~\cite{Nak16, Chen18, Muj21_q, Pen23} have also been demonstrated using a range of quantum technologies, including solid-state systems like quantum dots \cite{han08, los98}, superconducting qubits \cite{wen17, cla08} and trapped ions \cite{bla08, mon13}. Each of these platforms offers distinct advantages in terms of scalability and coherence time.

\begin{figure}[t]
\centering
\includegraphics[width=0.99\columnwidth]{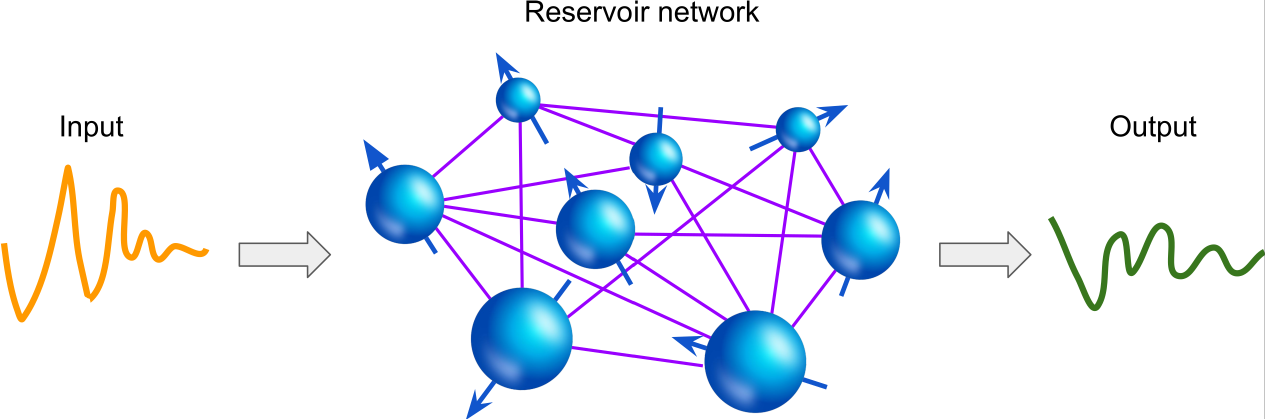}
\caption{Sketch of a spin-network-based QRC system. The system operates as follows. First, the classical input data are injected into the spin network, altering its quantum state. Then, the quantum dynamics of the network evolves under the influence of the input data. Finally, the network produces an output signal derived from certain observables of the system. Redrawn after Refs.~\cite{Fuj17, Muj21_q}.}
\label{Fig8}
\end{figure}

In the context of spin-network-based QRC, the reservoir consists of a network of quantum spins organised in a particular topology (Figure~\ref{Fig8}). The interactions between these spins govern the dynamics of the reservoir. These quantum spins can represent qubits or higher-dimensional quantum systems, providing a versatile computational platform for diverse information processing tasks \cite{Fuj17, Muj21_q}. Furthermore, the network topology and the Hamiltonian parameters can be fine-tuned to optimise the performance of the reservoir for specific applications. For example, by adjusting the strength and nature of the interactions between the spins, one can control the response of the reservoir to input signals and its ability to perform different computational tasks \cite{Fuj17, Tan19, Muj21_q}. Such a tuneability is crucial for tailoring the reservoir to the requirements of various applications, ranging from pattern recognition and time-series prediction to more sophisticated quantum information processing tasks \cite{Suz22}.

The concept of spin-network-based QRC system was introduced in Ref.~\cite{Nak16}, where the reservoir exploited the dynamical properties of a nuclear magnetic resonance spin-ensemble architecture, using a multiplexing technique to enhance its computational capabilities \cite{Nak19}. The so-created QRC system utilised the intrinsic properties of a network of qubits---the fundamental two-level quantum systems that reside in a two-dimensional complex Hilbert space \(\mathcal{H}^2\) \cite{Nie02}. The state of a qubit can be described as a linear combination of linearly independent bounded operators acting on \(\mathcal{H}^2\). These operators are formed from the tensor products of basic Pauli operators \(\{I, \sigma_z, \sigma_x, \sigma_y\}\) for each qubit. The state of a single qubit can be visually represented on the Bloch sphere, a three-dimensional space that represents the qubit's state vector \cite{Nie02}.

The dynamics of the reservoir layer is driven by the transverse-field Ising model \cite{Sin20} and it forms a high-dimensional space. The Hamiltonian of the transverse-field Ising model typically includes the terms that represent the interaction between neighbouring qubits and an external transverse magnetic field, and it is given by
\begin{equation}
H = J \sum_{i \neq j} \sigma_i^x \sigma_j^x + h \sum_i \sigma_i^z\,,
  \label{eq:RC2}
\end{equation}
where $N$ is the number of qubits, $h$ is an external magnetic field and $J_{ij}$ denotes the coupling strength between the qubits $i$ and $j$. The evolution of such closed quantum systems is described by the time-dependent Schr{\"o}dinger equation
\begin{equation}
 \rho(t) = e^{-iHt} \rho(0) e^{iHt}\,,
  \label{eq:RC3}
\end{equation}
where $\rho$ is the reservoir state, $H$ is the Hamiltonian of the system given by Eq.~\ref{eq:RC2} and $\hbar$ is the reduced Planck constant. 

The reservoir employs an amplitude encoding scheme to drive its quantum states. In this approach, the state of each qubit is initialised by a set of input signals $s_k$. The state of the qubit is reset at each time step to
\begin{equation}
|\psi_{s_k}\rangle = \sqrt{1 - s_k} |0\rangle + \sqrt{s_k} |1\rangle\,,
  \label{eq:RC4}
\end{equation}
represented by the density matrix $\rho_1 = |\psi_{s_k}\rangle \langle \psi_{s_k}| $. The updated reservoir state $\rho'_k$ at each time step is given by
\begin{equation}
\rho'_k = \rho_1 \otimes \text{tr}_1(\rho_{k-1})\,,
  \label{eq:RC5}
\end{equation}
where $ \text{tr}_1 $ denotes the partial trace over the state of the first qubit. 

There exist various strategies that can be used to implement the input protocol, including strong local dissipation followed by a quantum rotation gate \cite{san22} or projective measurements over the first qubit \cite{Lud50, von13}. Subsequently, the system evolves under its natural dynamics for a time interval \( \Delta t \) to process the information:
\begin{equation}
\rho_k = e^{-iH\Delta t} \rho'_k e^{iH\Delta t}\,.
  \label{eq:RC6}
\end{equation}
The output of the reservoir is obtained by measuring specific observables of the spin network. These observables are related to certain properties of the system, such as the magnetisation of individual spins and the correlations between spins at different locations within the network. The data obtained from these measurements serve as the output of the reservoir. This output can then be further processed using classical computational techniques such as linear regression approaches \cite{Jae04, Luk09}.

It is well-known that quantum systems exhibit physically rich dynamical regimes, involving the effects of localisation or thermalisation. An attempt to address the impact of these dynamical phases on a quantum computational reservoir was made in Refs.~\cite{Xia22, Pen23}. In those works, the Hamiltonian Eq.~(\ref{eq:RC3}) was extended with the aim to investigate the dynamical regimes \cite{Pen23}, where the condition for the homogeneous external field would be relaxed and a local magnetic field applied to each spin $\sigma_i^z$. It is also established that variations of the dynamical regimes of the reservoir influence its computational robustness \cite{Jae04, Luk09}. For instance, systems presenting localisation can provide quantum memories at finite temperature \cite{Ped15}, also improving the trainability of parameterised quantum Ising chains \cite{Alt10}. Yet, the thermal phase appears to be naturally adapted to the requirements of QRC since the performance of the reservoir increases at the thermalisation transition \cite{Pen23}. 

\subsection{Quantum Oscillator for Reservoir Computing}
Errors caused by decoherence and noise remain a significant challenge in the field of quantum information processing \cite{Hor22}. One potential solution of this challenge consists in adapting noise-resilient classical computing modalities to the quantum realm.

For example, the study conducted in Ref.~\cite{Gov21} proposed a continuously-variable QRC system based on a single nonlinear oscillator, demonstrating that such a quantum-mechanical system serves as a computational reservoir that outperforms its classical counterparts in both performance and reliability. In particular, the continuous-variable approach was implemented using a single nonlinear oscillator with the Kerr nonlinearity \cite{Gov21} (the state of this system can be described by continuous variables corresponding to its position $X$ and momentum $P$ quadrature). The advantage of using a continuous-variable system lies in its ability to provide a richer set of computational nodes due to the infinite-dimensional Hilbert space associated with these continuous variables, which, in turn, reduces costly repetitions required for accurate measurement of expectation values in discrete-variable quantum machine learning approaches \cite{Qiu19, Per24_2}.

The dynamics of a classical reservoir can be described by the differential equation 
\begin{equation}
\dot{a} = -iK(a - 2a^*) - \frac{\kappa}{2} a - i\alpha u(t) \,.
  \label{eq:RC7}
\end{equation}
In the case of a quantum reservoir, the Hamiltonian of the Kerr-nonlinear oscillator is
\begin{equation}
\hat{H}(t) = K \hat{a}^\dagger \hat{a} \hat{a}^\dagger \hat{a} + \alpha u(t)(\hat{a} + \hat{a}^\dagger) \,
  \label{eq:RC8}
\end{equation}
and its evolution is described by the Lindblad master equation \cite{Cam23}
\begin{equation}
\dot{\rho} = -i[\hat{H}(t), \rho] + \kappa D[\hat{a}] \rho \,.
  \label{eq:RC9}
\end{equation}

It has been established that a single-oscillator QRC system outperforms its classical counterpart in the task of sine wave phase estimation, primarily due to the nonlinearity of quantum measurements that is known to provide an intrinsic advantage by effectively performing nonlinear transformations on output data \cite{Gov21} (for a relevant discussion of nonlinear transformation in the context of the traditional reservoir computing see Refs.~\cite{Bol21, Gau21, Mak24_dynamics}). The QRC system also demonstrated smaller average root mean square (RMS) errors and greater reliability, with a lower sensitivity to parameter variations compared with a classical reservoir tasked with similar benchmarking problems. The performance of QRC enhanced with the increase in Hilbert space dimension, suggesting an increase in the number of degrees of freedom and boosting of its computational power \cite{Gov21}.

The impact of the size of the Hilbert space dimension on the computational power of QRC was investigated in Ref.~\cite{Kal22}. It was established that the use of higher-dimensional quantum oscillators helps improve the performance of the reservoir in both signal processing tasks and memory capacity benchmarking problems. Furthermore, quantum reservoirs demonstrated superior memory retention capabilities, additionally improving their performance at stronger nonlinearity and higher dimensionality \cite{Kal22}. 

Another implementation of QRC using coherently coupled quantum oscillators was suggested in Ref.~\cite{Dud23}. The approach proposed in that work aimed to address certain limitations of the existing quantum neural networks that rely on qubits, including the problem of limited connectivity between the artificial neurons. The authors of Ref.~\cite{Dud23} demonstrated that one can use coupled quantum oscillators to achieve a large number of densely connected neurons, thereby obtaining advantage over the traditional qubit-based approaches. 
\begin{figure}[t]
\centering
\includegraphics[width=0.8\columnwidth]{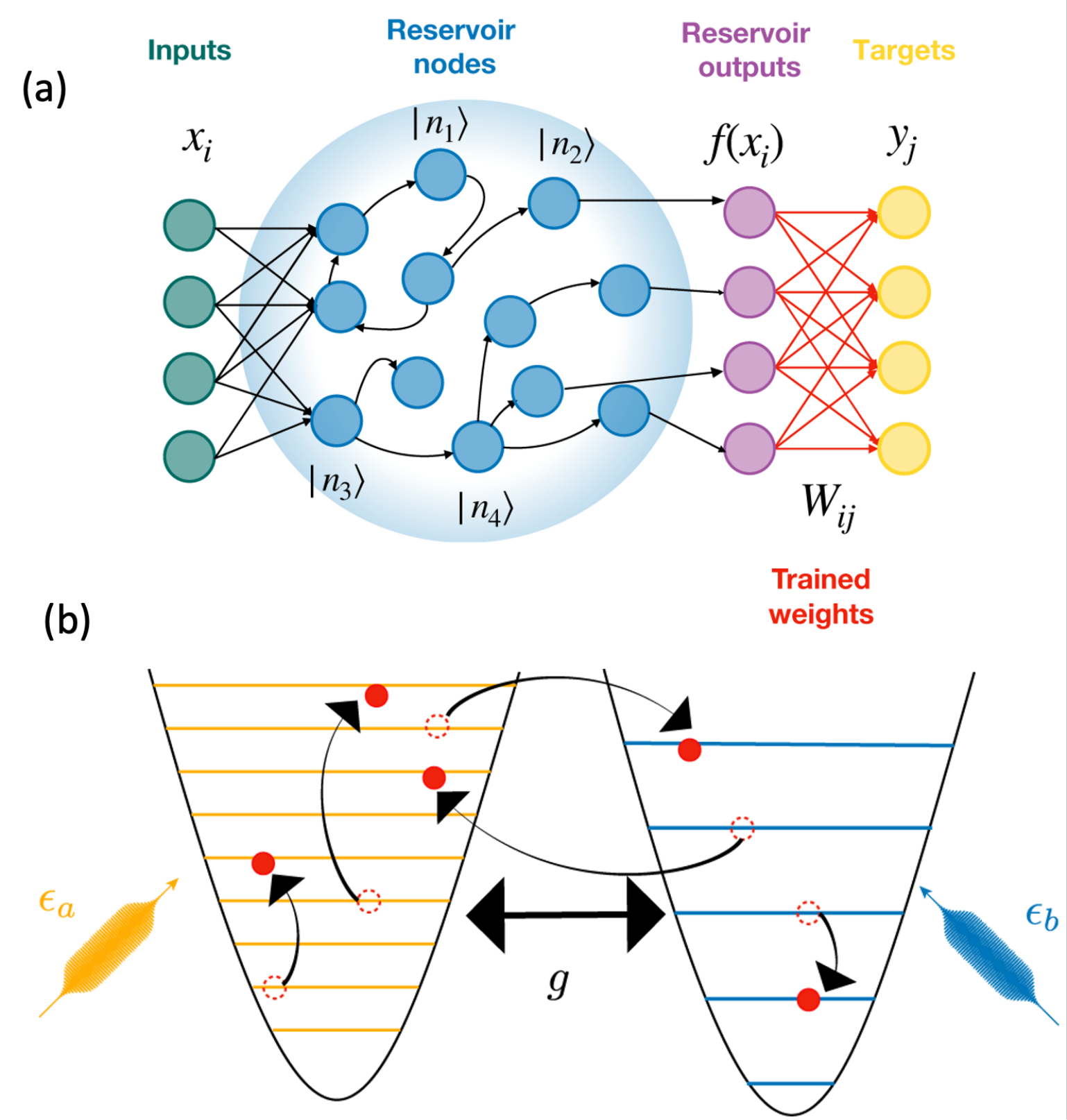}
\caption{{\bf(a)}~Sketch of a QRC using coherently coupled oscillators. The blue and yellow circles represent arbitrarily connected neurons of the reservoir. The black connections arrows denote the fixed weights but the red connections correspond to the trained weights. {\bf(b)}~Illustration of a system of coupled quantum oscillators driven at the frequencies $\omega_{a,b} $ and with the amplitudes $\epsilon_{a,b}$. Adapted from Ref.~\cite{Dud23} under the terms of a Creative Commons Attribution 4.0~International License.}
\label{Fig9}
\end{figure}

The system investigated in Ref.~\cite{Dud23} can be experimentally realised using superconducting circuits with resonators linked via parametric elements such as Josephson mixers or SNAILs (Superconducting Nonlinear Asymmetric Inductive eLements) employed as the source of tuneable nonlinearity \cite{Fra18}. Utilising Fock states as the functional neurons, the theoretical analysis carried out in Ref.~\cite{Dud23} revealed that a quantum reservoir with up to 81\,neurons based on two coupled oscillators (Figure~\ref{Fig9}) can achieve a 99\% accuracy rate on challenging benchmarking tasks. To put this result into perspective, a similar accuracy was achieved using at least 24\,classical oscillators.

The authors of Ref.~\cite{Dud23} also investigated the necessary coupling and dissipation conditions for optimal performance of the quantum reservoir proposed by them. They established that high dissipation rates lead to greater errors and reduced memory span for the neural network, which emphasised the need to employ high-quality-factor oscillators for computational problems that require extensive memory. Conversely, stronger coupling between the oscillators resulted in an increase in the reservoir's performance by enabling more significant data transformations between different basis states, which is crucial for effective learning. The operation in a strong coupling regime also facilitates a larger population of basis state neurons, thereby boosting computational capabilities. Overall, it was also demonstrated that the optimal performance of a quantum reservoir relies on finding a balance between the dissipation rate and coupling strength parameters, which maximises memory retention and data transformation efficiency.

However, despite the substantial progress made in the field of quantum computing, experimental realisation of quantum neural networks capable of handling real-world classification tasks remains elusive. In particular, such tasks require networks with millions of interconnected neurons to efficiently process complex data. However, as already mentioned above, traditional qubit-based approaches face connectivity limitations, making it challenging to achieve the necessary network density. To resolve this problem, it has been suggested that using 10\,coupled quantum oscillators could enable the creation of an analogue quantum neural network architectures with billions of neurons \cite{Dud23}. In the following subsection, we discuss alternative strategies aimed to additionally decrease the demand for the high number of artificial neurons in the reservoir. 

\subsection{Quantum Reservoir with Controlled-Measurement Dynamics}
While QRC is a promising platform for quantum information studies, as of time of writing the existing QRC models encounter execution time challenges due to the need for a repeated system preparation and measurement at each time step. To address this problem, improved QRC systems have been proposed relying on repeated quantum non-demolition measurements employed to generate time-series data and reduce the execution time \cite{Yas23}. The system described in Ref.~\cite{Yas23} was experimentally implemented using IBM’s quantum superconducting devices, and it demonstrated higher accuracy and smaller execution time compared with conventional QRC methods.

The concept of Temporal Information Processing Capacity was introduced into research practice to evaluate the computational abilities of QRC systems, highlighting the fact that an optimal measurement strength needs to be found to balance the retention and dissipation of information \cite{Yas23}. In the cited paper, the proposed QRC method was applied to a soft robot over 1000\,time steps, showcasing its practical utility. Initial experiments on a 120-qubit device were discussed, indicating the method's potential for scalability. 

Moreover, addressing the challenge of integrating quantum measurement while retaining processing memory and handling large Hilbert spaces, the authors of Ref.~\cite{Muj23} proposed additional measurement protocols aimed to achieve optimal performance for both memory and forecasting tasks. The three proposed measurement protocols for QRC are:
\begin{itemize}
    \item \textbf{Restarting Protocol (RSP)}: Repeats the entire experiment for each measurement, maintaining unperturbed dynamics but requiring significant resources.
    \item \textbf{Rewinding Protocol (RWP)}: Restarts dynamics from a recent past state (washout time $\tau_{wo}$), optimising resources compared to RSP.
    \item \textbf{Online Protocol (OLP)}: Uses weak measurements to continuously monitor the system, preserving memory with less back-action but introducing more noise.
\end{itemize}
The efficiency of these measurement protocols was evaluated by tasking a QRC system to solve a number of standard benchmarking problems. It was established that RSP and RWP require strong measurements to minimise statistical errors, whereas OLP with weak measurements effectively preserves the state of the reservoir while providing sufficient information for time-series processing \cite{Muj23}.

Nevertheless, the ambiguity of quantum mechanics and quantum measurements continues to inspire research on the topic of quantum reservoir computing. A novel quantum RC architecture that exploits the dynamics of an atom trapped in a cavity was recently proposed in Ref.~\cite{Abb24}. One prominent feature of the proposed system is a coherent driving of the atom at a certain driving rate with the possibility to observe transitions between quantum states and effective ``freezing'' of the quantum evolution of the system. Frequent observations of the atom eigenstates prevent the system from undergoing significant changes, a phenomenon known as the Zeno effect \cite{Harr17}. Conversely, less frequent probing of the system states enables the system to undergo Rabi oscillations \cite{Rai12, Lew23}. Using these properties, the rate at which the atom is driven becomes optimised, making the quantum dynamics of the reservoir suitable for undertaking diverse classification and prediction tasks. This approach also enables controlling and stabilising quantum states during a computation, which is beneficial for such practical applications as mitigation of decoherence \cite{Yas16}, quantum information processing \cite{Ale09}, correction of quantum error and stabilisation of quantum states \cite{Paz12, Lew23, Bur23}.
\begin{figure}[t]
\centering
\includegraphics[width=0.99\columnwidth]{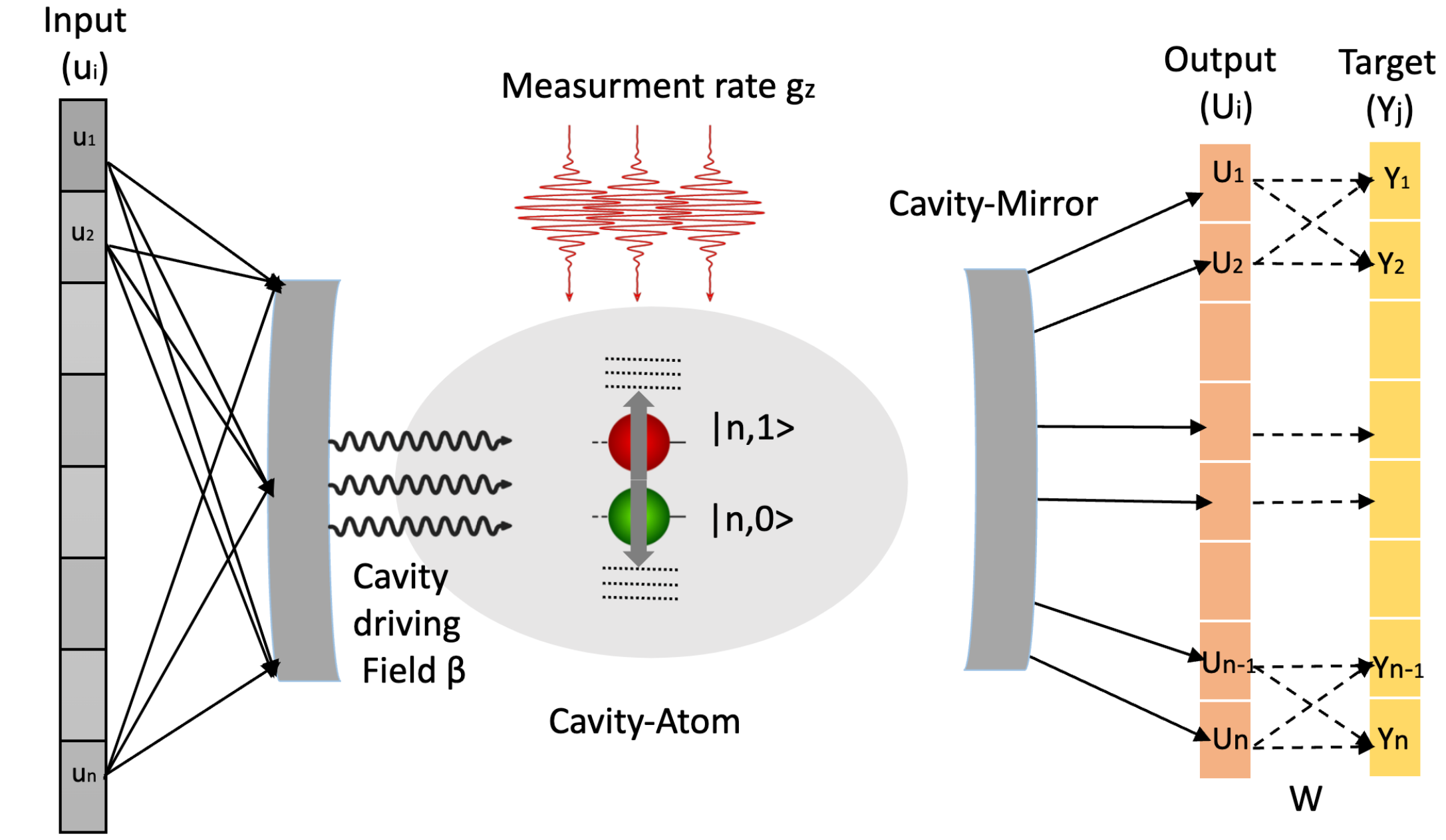}
\caption{Sketch of the quantum reservoir computing system employing measurement-controlled dynamics. The input, output and target data are defined similarly to the notation used in Figure~\ref{Fig1_1}. The quantum reservoir is constructed using the Fock states $ |n, \sigma\rangle $ representing the quantum states within the cavity. Input data points are encoded as signals modulating the driving amplitude $ \beta(t) $. The output of the reservoir is determined by the expectation values $ P(n, \sigma) $ of the occupancy of the basis states.}
\label{Fig10}
\end{figure}

The theoretical QRC system proposed in Ref.~\cite{Abb24} exploits the dynamics of a probed atom in a cavity (Figure~\ref{Fig10}). Coherently driving the atom and inducing measurement-controlled quantum evolution, the authors of Ref.~\cite{Abb24} demonstrated that the QRC system can exhibit fast and reliable forecasting capabilities using fewer artificial neurons compared with the traditional RC algorithms. The computational tests of the system also showcased its potential to be employed in error-tolerant applications, particularly in conditions of limited computational and energy resources. 

The model developed in Ref.~\cite{Abb24} is constructed around the interaction of a two-level atom with a quantised electromagnetic field in a cavity, enabling real-time observations and control of the quantum state of the atom. The interaction between the atom and the cavity is governed by the Hamiltonian
\begin{eqnarray}
\hat{H}_i = g a^\dagger a \sigma_{-} \sigma_{+}\,,
\label{eq:RC10}
\end{eqnarray}
where $g$ represents the strength of the atom-cavity coupling, $a$ is the cavity annihilation operator and $\sigma^-$ and $\sigma^+$ are the lowering and raising operators for the atom, respectively. The cavity is coherently driven by an external signal, which is modelled by the Hamiltonian
\begin{eqnarray}
\hat{H}_c = -i\beta(a^\dagger - a)\,,
\label{eq:RC11}
\end{eqnarray}
but state of the atom undergoes continuous monitoring by means of coherent measurement processes, which accounted for by the Hamiltonian
\begin{eqnarray}
\hat{H}_z = g_z (\sigma_{+} + \sigma_{-})\,,
\label{eq:RC12}
\end{eqnarray}
where $g_z$ denotes the amplitude of the coherent atomic driving that controls the frequency at which the state of the atom is measured. The dynamics of the entire system are described by the stochastic master equation
\begin{eqnarray}
\dot{\rho }=-i[\hat{H},\rho ]+\hat{C}\rho {\hat{C}}^{{\dagger} }-\frac{1}{2}{\hat{C}}^{{\dagger} }\hat{C}\rho -\frac{1}{2}\rho {\hat{C}}^{{\dagger} }\hat{C}\,,
\label{eq:RC13}
\end{eqnarray}
where $\rho$ is the density matrix and $ \mathcal{C} = \sqrt{\kappa} a $ is the collapse operator associated with cavity decay.

Incorporating these components and equations, the model of the proposed QRC system revealed the system's capability to efficiently process input data, adapt its dynamics through coherent driving and measurement and make accurate predictions while using low computational and energy resources compared with the traditional RC algorithms.

\section{Conclusions and Outlook}
Thus, in this article we have reviewed the recent advances in the fields of classical and quantum reservoir computing systems. Although the potential practical applications considered by us in this work are mostly limited to onboard AI systems of autonomous vehicles, the neuromorphic computational reservoirs discussed throughout the main text are expected to find numerous applications in the broad area of science and technology.  

Indeed, at present the world of AI has been enabled by well-established technologies that are about to reach their fundamental operation limits and already consume the power comparable with the annual electricity generation by a medium-size country. Subsequently, since the computational power required for sustaining the demand for novel AI units is doubling approximately every three months, humankind will soon face the problem of finding a balance between the progression of AI with the imperatives of sustainable development.

Of course, the problem is not exclusively limited to managing the rise of AI and transitioning to green energy. In fact, since several billions of self-driving car, drones and autonomous robotic systems are estimated to enter service by 2050, each of those vehicles will require a source of energy exclusively dedicated to powering their onboard AI. Subsequently, since no current technology is able to satisfy such a high demand for lightweight and high-capacity energy sources, humans will need to invent alternative, low power consuming computational technologies.

In his famous talk ``Plenty of Room at the Bottom'' addressed to the American Physical Society in Pasadena on December 1959 \cite{feynman_nano}, Richard P.~Feynman suggested exploring the immense possibilities enabled by miniaturisation of devices, thus heralding the coming era of nanotechnologies and quantum computing. A large section of this present review article has been dedicated to computational approaches that directly benefit from the suggestions made by Feynman in that talk. Yet, standing on the shoulders of giants and rephrasing Feynman, in this article we invite researchers to enter a new field of AI, pointing out that that there is plenty room and rich physics around autonomous vehicles to employ them as a means of energy-efficient computing.   

\vspace{6pt} 

\authorcontributions{The idea to write this review article belongs to all authors. I.S.M.~wrote Sections~1-3, 5 and 6. Section 4 was written by H.A.G and I.S.M. Section~7 was written by A.H.A. All authors discussed the manuscript and approved the submitted version of the paper.}

\funding{This research received no external funding}

\institutionalreview{Not applicable.}

\informedconsent{Not applicable.}

\dataavailability{This article has no additional data.} 


\conflictsofinterest{The authors declare no conflicts of interest.} 



\abbreviations{Abbreviations}{
The following abbreviations are used in this manuscript:\\

\noindent 
\begin{tabular}{@{}ll}
AI & artificial intelligence\\
LIDAR & Light Detection and Ranging\\
NMR & nuclear magnetic resonance\\
OLP & online protocol\\
QNP & quantum neuromorphic processor\\
QRC & quantum reservoir computing\\
RC & reservoir computing\\
ROV & remotely operated vehicle\\
RSP & restarting protocol\\
RWP & rewinding protocol\\
SNAILs & Superconducting Nonlinear Asymmetric Inductive eLements\\
UAV & unmanned aerial vehicle\\
UGV & unmanned ground vehicles
\end{tabular}
}

\begin{adjustwidth}{-\extralength}{0cm}

\reftitle{References}


\bibliography{refs}

\end{adjustwidth}
\end{document}